\documentstyle[12pt,epsf]{article}
\begin{document}

\begin{center}
{\LARGE{\bf
Projectile $\Delta$ and target-Roper excitation in the $p (d, d')X$ 
reaction}}
\end{center}

\vspace{0.5cm}

\centerline{S. Hirenzaki$^{1}$, E.~Oset$^{2}$, C.~Djalali$^{3,4}$, 
M.~Morlet$^{4}$}

\vskip 0.5 true cm

\centerline{{$^1$}{\it Department of Physics, Nara Women's University, 
Nara 630-8506, Japan}}
\centerline{{$^2$}{\it Departamento de F\'{\i}sica Te\'orica and IFIC}}
\centerline{{\it Centro Mixto Universidad de Valencia - CSIC,}}
\centerline{{\it 46100 Burjassot (Valencia), Spain}}
\centerline{{$^3$}{\it Physics Department, University of South Carolina, 
Columbia, SC 29208}}
\centerline{{$^4$}{\it CNRS-IN2P3, I. P. N., 91406-Orsay, France}}

\vskip 1.0 true cm

\begin{abstract}
In this paper we compare a model that contains the mechanisms of $\Delta$ 
excitation in the projectile and Roper excitation in the target with 
experimental data from two $(d, d')$ experiments on a proton target. The 
agreement of the theory with the experiment is fair for the data 
taken at T$_d$ = 2.3~GeV. The $\Delta$ excitation in the projectile is 
predicted close to the observed energy with the correct width. The theory, 
however, underpredicts by about 40\% the cross sections measured at T$_d$ 
= 1.6~GeV at angles where the cross section has fallen by about two orders 
of magnitude. The analysis done here allows to extract an approximate 
strength for 
the excitation of the Roper [N$^*$(1440)] excitation and a 
qualitative agreement with the theoretical predictions is also found.
\end{abstract}

\vskip 1.0 true cm

\section{Introduction}

The $(\alpha, \alpha')$ reaction on a proton target measured at Saclay 
\cite{1} has been instrumental in setting the question of the mechanism of 
$\Delta$ excitation in the projectile (DEP), which was introduced in 
\cite{2} in order to describe the $(^{3}He, t)$ reaction in p and d 
targets \cite{3}. That mechanism plays a negligible role in the $(^{3}He, 
t)$ reaction on proton targets but is quite important in the same reaction 
on neutron targets and was predicted to be dominant in the $(^{3}He, 
^{3}He)$ reaction on proton and neutron targets \cite{4}. Prior to the 
$(\alpha, \alpha')$ experiment the relevance of the DEP mechanism was a 
subject of debate \cite{5,6}, particularly because of the small strength 
of this mechanism in the $(^{3}He,t)$ reaction on proton targets, which 
allowed interpretations omitting it \cite{7,8}. The $(\alpha, \alpha')$ 
reaction on a proton target is ideal to isolate the DEP mechanism since, 
because of isospin, the $\Delta$ excitation on the proton target is 
forbidden. This allowed to test the ideas introduced in Ref~\cite{2} and 
indeed the large peak in the experiment \cite{1} corresponding to DEP was 
well reproduced\cite{9}.

In addition to the issues discussed above, the ($\alpha,\alpha')$ 
experiment \cite{1} observed a smaller peak at higher excitation energies 
which was attributed to the Roper excitation. This mode of excitation of 
the Roper is novel, since it involves an isoscalar source, and can be 
relevant in determining the strength of three body forces \cite{10} and 
providing new tools for the comprehension of the $N N \rightarrow N N \pi 
\pi$ and related reactions \cite{11}.

The strength of the isoscalar $N N \rightarrow N N^*$ transition was 
determined empirically from the experimental data in \cite{12}, where the 
model of \cite{9} for DEP was used and the interference of the two 
mechanisms was also considered. The analysis proved consistent with the 
present knowledge of position, decay width and partial decay widths of the 
Roper and helped to narrow the experimental uncertainties on these 
magnitudes.

The issue of strong isoscalar excitation of nucleon resonances has 
captured more attention after the first inclusive $(\vec {d}, d')$ data 
were obtained~\cite{fewbody}. The T$_{20}$ data obtained at Dubna on the 
proton and $^{12}$C, are given in  Ref~\cite{PLB361} and the final data 
tables are published in Ref~\cite{JINR}.  In Ref~\cite{13} polarization 
observables for the $(\vec {d}, d')$ reaction on proton targets are 
discussed bringing new information on electromagnetic form factors of the 
deuteron and on mechanisms for strong excitation of nucleonic resonances. 
The description done in Ref~\cite{12} was extended to higher energies of 
around 10-15 GeV \cite{14}, showing that the magnitude of the Roper 
excitation can be increased by about one order of magnitude and the 
relative strength of the Roper signal to the one of the DEP mechanism 
becomes of the order of unity, much bigger than in the $(\alpha, \alpha'$) 
experiment \cite{1} where it is about 1/4. In Ref~\cite{15} polarization 
observables in the $(\vec{p}, \vec{p} \, ')$ on a $^4$He target are 
studied with its view towards possible experiments to be carried out at 
the Indiana Cyclotron.

The program of nucleon resonance excitation using baryonic interactions is 
thus catching up, and certainly will bring complementary information to 
the one obtained with electromagnetic probes or meson induced excitation.

The present work, DEP and Roper excitation on the $(d, d')$ reaction on 
proton targets, should be considered as a complement to the one of the 
$(\alpha, \alpha')$ reaction \cite{12}. The fact that the deuteron has an 
isospin $I = 0$ makes the two works similar since $\Delta$ excitation on 
the proton target is forbidden in both cases and only DEP and Roper 
excitation are allowed in the region which we study. However, the fact 
that the deuteron has a total spin $J = 1$ induces some differences with 
respect to the $(\alpha, \alpha')$ reaction and sets different constraints 
on the theoretical models. With new experiments on this issue coming, the 
need to have reliable theoretical models to extract the relevant 
information becomes apparent, and in this sense our present work is a 
valuable one. We have taken advantage of the existence of experimental 
information from Saclay measurements of the $(d, d')$ reaction and we 
present here a paper where the theoretical ideas are exposed, the data are 
presented and a discussion is made from comparison of theory and 
experiment.

Early work on the present reaction at different kinematics is done in Ref~ 
\cite{banaigs73}.  We will compare the theoretical results of the model 
with recent (d,d') measurements done at Saclay \cite{rapintern} and older 
(d,d') measurements \cite{17} done at lower energies and larger angles.

\section{Formulation}

In this section we consider a theoretical model of the $(d, d')$ reaction 
on the proton target. We include two processes, $\Delta$ excitation in the 
projectile and Roper excitation in the target, which are shown in Fig. 1 
and are the dominant processes in this energy region \cite{12}. We include 
both the $\pi N$ and $2 \pi N$ decay modes of the Roper resonance. Since 
we need to take care of the interference between the projectile $\Delta$ 
process and the target Roper process decaying into $\pi N$, we treat the 
Roper $\rightarrow \pi N$ and Roper $\rightarrow 2 \pi N$ processes 
separately. We take the same model which was used to analyze the $(\alpha, 
\alpha')$ reaction at $4.2$ $GeV$ and use the same values for all 
parameters \cite{12}.

The cross section for the $1 \pi$ decay $\Delta$ and Roper processes is 
given by,

$$
\frac{d^2 \sigma}{d E_{d'} d \Omega_{d'}} = \frac{p_{d}'}{(2 \pi)^5} \;
\frac{M^2_d M^2}{\lambda^{1/2} (s, M^2, M^2_d)} \;
\int d^3 p_{\pi} \; \frac{1}{E'_N \omega_{\pi}}
$$

\begin{equation}
\times \bar{\Sigma} \Sigma | T^{1 \pi} |^2 \delta (E_d + E_N - E_d' - E'_N 
- \omega_{\pi})
\end{equation}

\noindent
where $\lambda$ (...) is the K\"allen function and s the Mandelstam 
variable for the initial p-d system, and momentum conservation, $\vec{p}_d 
+ \vec{p}_N = \vec{p}_{d}\, ' + \vec{p}\, '_{N} + \vec{p}_{\pi}$ is already 
implied. The projectile $\Delta$ mechanism ($T_{\Delta}$) leads to a $\pi 
N$ through the decay of the $\Delta$. Part of the Roper excitation 
mechanism leads to the same final state through the decay of the $N^*$ 
into $\pi N$.  We call this latter piece $T^{1 \pi}_*$.  Hence the sum of 
the two mechanisms leading to $\pi N$ is given by;

\begin{equation}
T^{1 \pi} = T_{\Delta} + T^{1 \pi}_*
\end{equation}

The nucleon and deuteron spin sum and average of each $|T|^2$ and plus the 
interference term can be written as,

\begin{eqnarray}
\bar{\Sigma} \Sigma |T_{\Delta}|^2 & = & \frac{16}{27} F^2_d \;
(\frac{f^*}{\mu})^4 \; (\frac{f}{\mu})^2 \; |G_{\Delta}|^2 \;
(\frac{- q^2}{\vec{q}^2}) \nonumber\\
& \times & [(V^2_{l'} + 5 V^2_{t'}) \vec{p}\,^2_{\Delta} + 3 (V^2_{l'} -
V^2_{t'})
(\vec{p}_{\Delta} \cdot \hat{q})^2]
\end{eqnarray}

\begin{equation}
\bar{\Sigma} \Sigma |T^{1 \pi}_*|^2 = 12 F^2_d (\frac{f'}{\mu})^2 \;
g^2_{\sigma N N} g^2_{\sigma N N^*} |G_*|^2 |D_{\sigma} F_{\sigma}^2|^2
\vec{p}\, ^2_*
\end{equation}

\begin{eqnarray}
 & & \bar{\Sigma} \Sigma ( T^{1 \pi *}_* T_{\Delta}  +  T^*_{\Delta} T^{1
\pi}_*)
\nonumber\\
& = & 2 Re \left\{ \frac{16}{3} F^2_d \frac{f'}{\mu} \;
(\frac{f^*}{\mu})^2 \;
\frac{f}{\mu} \; g_{\sigma N N} \; g_{\sigma N N^*} \; D^*_{\sigma} \;
G^*_* \; G_{\Delta} \; F^2_{\sigma} \; \right.  \nonumber\\
& \times & [V_{l'} (\vec{p}_* \cdot \hat{q}) \; (\vec{p}_{\Delta} \cdot
\hat{q})
+ V_{t'} \; (\vec{p}_* \cdot \vec{p}_{\Delta} - (\vec{p}_* \cdot \hat{q})
(\vec{p}_{\Delta} \cdot \hat{q}))]
\left. \right\} \; \sqrt{\frac{- q^2}{\vec{q}\, ^{2}}}
\end{eqnarray}

\noindent
where $G_{\Delta}$ and $G_*$ are the propagators of the $\Delta$ and Roper 
resonances, $D_{\sigma}$ the propagator of the $\sigma$ meson, 
$F_{\sigma}$ the $\sigma N N$ vertex form factor. The momenta $\vec{p}_*, 
\vec{p}_{\Delta}$, and $\vec{q}$ are the pion momenta in the Roper rest 
frame, pion momentum in the $\Delta$ rest frame, and momentum transfer 
between the nucleons, respectively. $V_{l'}$ and $V_{t'}$ stand for the 
longitudinal and transverse parts of $N N \rightarrow N \Delta$ effective 
interaction which includes $\pi, \, \rho$ and $g'$ contributions, where 
$g'$ is the Landau-Migdal parameter which is meant to account for short 
range corrections to the $\pi$ and $\rho$ exchange. The $f' s$ and $g' s$ 
are coupling constants. All details, including parameter values, are shown 
in ref. \cite{12}.

The function $F_d$ is the deuteron form factor defined as

\begin{equation}
F_d (\vec{k}) = \int d^3 r \varphi^* (\vec{r}) \;
e^{i \vec{k} \cdot \frac{\vec{r}}{2}} \; \varphi (\vec{r})
\end{equation}

\noindent
where $\varphi (r)$ is the relative wave function of the deuteron obtained 
from the Bonn potential \cite{16}. The momentum transfer of the deuteron 
is denoted by $\vec{k} = \vec{p}_d - \vec{p}_{d'}$ taken in the initial 
deuteron rest frame. We have included only the s-wave part of the deuteron 
wave function for simplicity. The contribution from the target Roper 
process decaying into $2 \pi N$ is calculated separately as,

\begin{equation}
\frac{d^2 \sigma}{d E_{d'} d \Omega_{d'}} = \frac{p_{d}'}{(2 \pi)^3} \;
\frac{2 M^2_{d} M}{\lambda^{1/2} (s, M^2, M^2_d)} \; \bar{\Sigma} \Sigma \;
|T^{\pi \pi}|^2 \; |G_*|^2 \; \Gamma^{\pi \pi}_*
\end{equation}

\noindent
with

\begin{equation}
\bar{\Sigma} \Sigma \; |T^{\pi \pi}|^2 = 4 F^2_d \; g^2_{\sigma N N} \;
g^2_{\sigma N N^*} \; |D_{\sigma} F^2_{\sigma}|^2
\end{equation}

\noindent
using the partial decay width, $\Gamma^{\pi \pi}_*$, whose explicit form 
is shown in the Appendix of ref. \cite{12}. This contribution is added to 
the $1 \pi$ contributions incoherently.

\section{Discussion of Experimental results}

We compare our theoretical results with two independent experimental data 
sets. One has been obtained recently at $T_d= 2.3 \, GeV$ at 
Saturne~\cite{rapintern} and another was measured some years ago at $T_d= 
1.6 \, GeV$ \cite{17}.

First, we consider the new data which were obtained during a short run at 
Saturne with a deuteron beam of 2.3~GeV.  The deuterons were directed onto 
a 4~cm thick liquid hydrogen target with thin Ti windows (15~$\mu m$). 
After going through a 40~cm thick lead collimator, the scattered deuterons 
were momentum analyzed at very small angles using the SPES4 
spectrometer~\cite{spes4} with a momentum acceptance of $\pm$~3\% and a 
resolution of $\approx$~10$^{-3}$. In the focal plane of the spectrometer, 
the scattered deuterons were detected using the three front wire chambers 
of the extended vector polarimeter POMME~\cite{pomme}.

In this experiment, special care was taken to minimize all possible 
experimental backgrounds. A missing mass spectrum was measured at 1.1$^o$ 
in five different momentum bites, respectively centered on 2.6, 2.8 3.0, 
3.2 and 3.45 GeV/c, covering an excitation energy region up to 600~MeV. 
The spectrometer acceptance is 15\% of the central momentum leading to 
smaller momentum bites for higher excitation energies. In order to get the 
most out of the limited beam time in terms of number of counts in each 
bite and range of coverage in excitation energy, the momentum bites were 
set as follows: two overlapping bites around 200~MeV of excitation energy, 
where the excitation of the $\Delta$ resonance in the projectile is 
expected, and, three non overlapping bites spanning the excitation energy 
range from 300 to 600~MeV, where the wide Roper resonance is expected (the 
kinematical limit is at 680~MeV of excitation energy). The target was 
located outside of the magnetic field of the spectrometer so the usual 
SPES4 corrections for correlations between scattering angle and scattered 
momentum were not necessary.

The cross section spectrum was binned in 10~MeV steps of excitation 
energy. For each setting of the spectrometer, empty target measurements 
were taken for background subtraction. The ratio of full to empty target 
was in the range of 8 to 10 dropping to 2 for the lowest setting of the 
spectrometer (corresponding to high excitation energies). We could not get 
clean measurement at momenta smaller than 2.4~GeV/c because of the large 
background due to rescattering through the lead collimator. When present, 
this background shows a strong angular dependence; taking into account the 
shape of the collimator, different cuts on angular acceptance lead to 
substantial changes in the shape and slope of the spectrum. For all 
momentum bites shown in Fig.~2, the off-line analysis of full versus empty 
target spectra and software cuts on angular acceptances showed no changes 
in the shape of the spectrum; only overall scaling of the spectrum 
consistent with the changes of the solid angle, were observed. This 
extensive off-line analysis led to the conclusion that no significant 
experimental background was present for these momentum bites.  Absolute 
cross sections were determined using monitors calibrated with the Carbon 
activation method~\cite{norm}. This is the well tested standard method of 
normalization used at Saturne.

The measured missing mass spectrum shown in Fig.~2 is dominated by a large 
structure centered around 200~MeV. This structure has been identified as 
the excitation of the Delta resonance in the incoming deuteron. The cross 
section drops sharply between 200 and 300~MeV of excitation energy. 
Between 300 and 600~MeV, there seems to be an excess of cross section 
above the high energy tail of the $\Delta$ resonance. There seems to be a 
discontinuity in the measured cross section around 500~MeV of excitation 
energy. This corresponds to the highest excitation energy bite that we can 
cleanly measure and could be affected by the subtraction of a relatively 
large empty target contribution.

In Fig. 2, we compare our theoretical results (including all reaction 
mechanisms described in section 2) with these experimental data. We have 
plotted on the same figures the different curves corresponding to the 
excitation of the $\Delta$ alone, the excitation of the Roper alone and 
the total cross section taking into account the interference between the 
two. For the excitation of the Roper, the contributions for the $\pi$N and 
2$\pi$N channels are also shown. We see that the projectile $\Delta$ 
excitation makes a large contribution to the cross section around 200~MeV. 
The calculation correctly reproduces the excitation energy of the $\Delta$ 
and its width, however it over-predicts the cross section at the maximum 
by at least 20\%. The excitation of the $\Delta$ in the projectile cannot 
account for the observed cross section between 300 and 500~MeV. This range 
of excitation energy is where we are expecting the Roper resonance to be 
and the calculation indeed predicts this excess of cross section to be due 
to the excitation of the Roper resonance. The calculation again 
overpredicts the measured cross section in this region.

In order to extract the excitation of the Roper in the 
200 to 600~MeV region, we assume that the shape of the calculated DEP is 
correct and we normalize it to the experimental spectrum at the maximum of 
the $\Delta$. This leads to an overall normalization factor of 0.85. We 
then subtract from the measured data the calculated differential cross 
section for the excitation of the $\Delta$ and the interference between 
Roper excitation and $\Delta$ excitation in the projectile, as described 
in eq. (5), also multiplied by the same normalization factor 0.85. What is 
left should mainly correspond to the excitation of the Roper resonance 
plus some physical continuum. In Fig.~3 we plot the measured data points, 
the predicted calculations normalized by 0.85 and the excess of cross 
section left once the DEP and interference contributions are subtracted. 
The 
excess of cross section has a maximum around 400~MeV, a width at half 
maximum of about 230~MeV, and an asymmetric shape with a long low energy 
tail. The calculation of the Roper contribution, normalized by the factor 
0.85, agrees qualitatively in shape and strength with the experimental 
cross section left once we have subtracted the DEP and the interference. 
Only the theoretical peak is shifted to higher excitation energy by about 
25 MeV. The total experimental excess cross section, up to 540~MeV, is 
32~$\pm$~8~mb/sr to be compared to the predicted cross section of 28 mb/sr 
for the Roper resonance. As mentioned earlier, this 
experimental cross section should be in principle an upper limit since no underlying continuum 
corresponding to other physical processes has been subtracted.
However, in \cite{12} other possible mechanisms were studied, and they 
were found to be small, leaving only the $\Delta$ excitation in the 
projectile and Roper excitation in the target as responsible for the 
reaction cross section in the energy region studied here.  
According to this, the signal obtained here for the Roper excitation,  
within experimental and theoretical uncertainties, should be rather fair.  

An empirical way to subtract the DEP contribution was done in 
Ref~\cite{rapintern}, and this is shown in Fig.~4. The shape of the 
excitation of the $\Delta$ in the projectile is taken from the 
measurements of Baldini et al~\cite{17}. The assumption being that at 
T$_d$=1.6~GeV, the measured spectrum is mainly dominated by the DEP 
mechanism and therefore its shape is a good empirical shape for this 
excitation. This shape is normalized to the data obtained at 
T$_d$=2.3~GeV, at the maximum of the $\Delta$ resonance. Once this 
empirical DEP contribution and also the interference contribution 
evaluated as in the previous case are subtracted, a wide structure is 
left, centered at 350~MeV with a width at half maximum of 230~MeV. The 
total cross section in this structure (up to 540~MeV of excitation energy) 
is of the order of 34~$\pm$~8~mb/sr. This value is in agreement with our 
previous determination of the excess cross section. The shape of the 
excess cross section is more symmetric than on the previous case and this 
is possibly due to the fact that the empirical spectrum taken from 
Ref.~\cite{17} contains already some contribution from the excitation of 
the Roper.

We have also compared our theoretical results with data at $1.6 \, GeV$ 
\cite{17}. The data measured at 6.59$^o$ and 8.05$^o$ are respectively 
compared to our predictions in figures 5 and 6.

 Our calculated results reproduce the overall shape of the spectrum well. 
However, the theoretical results are about $30 \%$ smaller than the data 
at 6.59$^o$ and about $40 \%$ at 8.05$^o$. This discrepancy has to be 
looked, however, in the proper perspective. Indeed, the angles where the 
cross sections are measured in \cite{17} are $\theta \geq 6.6^0$. At the 
smallest angle the cross section has fallen by a factor 30 from the 
forward direction. This fall down is mostly due to the deuteron form 
factor which involves large momentum transfers. In this case our neglect 
of the d-wave in the deuteron is not justified. This, and other 
approximations could explain these discrepancies. In view of the fact that 
they  represent only about $1 \%$ of the integrated cross section, we pay 
no further attention to these discrepancies, but the qualitative agreement 
found gives also partial support to the model.

In what follows we would like to make some estimates of the theoretical 
uncertainties in the present analysis of the data. One of the sources of 
uncertainty is our neglect of the d-wave in the deuteron wave function. 
The 
other one is the possible effect of Fermi motion in the deuteron . These 
two factors could change the shape on the $\Delta$ excitation strength and hence lead
 to uncertainties in the Roper excitation after the DEP strength and
 interference are subtracted from the data. 
 
   We begin by the effect of Fermi motion. The deuteron Fermi motion is 
   considered here in the same way as done in \cite{9}. It affects the $\Delta$
   propagator which enters the evaluation of the DEP mechanism. This propagator 
 is given by
 
 \begin{equation}
 G_{\Delta}(s)=\frac{1}{\sqrt{s} - M_{\Delta} + \frac{i}{2}  
 \Gamma_{\Delta} (s)} 
\end{equation}

\noindent 
 where the variable s is taken as 
 
\begin{equation}
s = (q^0 + M)^2 - \left( \frac{\vec{q} + \vec{p}_{\pi}}{2} \right)^2 ,
\end{equation}
 
 \noindent
 where $q$ and $p_\pi$ are the  momenta of the exchanged meson, and the emitted
 pion, respectively , taken in the frame of reference where the deuteron is at
 rest.  In this approximation the momentum transfer is shared equally by the
 initial and final nucleon in the deuteron.  The fairness of this 
 approximation to account for Fermi motion of the nucleus was well established in
 Refs \cite{boffi,carrasco} in the study of coherent pion
 photoproduction with similar momentum transfers as here. However, in order to
 see the effects of Fermi motion and have a feeling for possible
 uncertainties from this source we have conducted new calculations in which in
 eq. (10) we assume the initial momentum of the struck nucleon of the
 deuteron to be zero. This replaces $\frac{\vec{q}+\vec{p_\pi}}{2}$ in that
 equation by $\vec{q}$.

   We can see the results of the new calculation in Fig.7. We can see that the
   strength of the $\Delta$ excitation is increased by about 20 \%.  The prescription
 followed here to account for Fermi motion was found in Refs 
 \cite{boffi,carrasco} 
 to be rather accurate, but even then we see that the effects of
 ignoring it altogether does not bring drastic changes in the cross section. The
 possible uncertainties from this source are further minimized if we normalize
 the theoretical results to the experimental cross section, as we have done in
 the analysis of this work. Indeed, if we do so we obtain the results shown in
 Fig. 8, which are shown superposed to those of Fig.3. As we can see there, the
 differences found between neglecting the Fermi motion , or taking it according
 to our prescription, are very small up to 500 MeV of excitation energy, once 
 the normalization of the cross section around the peak of the delta resonance 
 is done.

    As for the d-wave of the deuteron we have proceeded as follows:
  In the evaluation of the deuteron form factor of eq. (6) we have taken only
  the s-wave of the deuteron so far. Inclusion of the d-wave into our scheme
  would lead to two parts. One with the same structure as we have, which goes
  with the $j_0(\frac{kr}{2})$ component of the exponential, but
  substituting $u^2$ by $u^2+w^2$ in eq. (6) ($u$ and $w$ are the s and d-wave
  parts of the deuteron wave function, respectively), and another one which 
   goes with the  $j_2(\frac{kr}{2})$ component of the exponential. Detailed
   evaluations of these two parts in the deuteron form factor can be seen in
   \cite{Glendenning} and there we see that up to 500 MeV/c the $j_2$ part of
   the form factor contributes less than 10\%. On the other hand the difference
   between the $j_0$ component evaluated with just the s-wave (normalized to
   unity) on the s plus d-wave parts of the deuteron wave function are smaller
   than 4\% up to this momentum.  In our case 500 MeV/c momentum transfer
   corresponds to an excitation energy of around 550 MeV in Fig. 2, just the
   tail of the distribution beyond the Roper excitation region which we have
   studied here. On the other hand in the case of Fig. 5 , 500 MeV/c would
   appear at $p_d$ around 2.25 GeV/c and in Fig. 6 one has already 500 MeV/c
   momentum transfer around the peak of the distribution. Hence, our neglect
   of the d-wave part of the wave function would induce more uncertainties, in
   the line we discussed above when we discussed the Baldini's data. 
   
     Altogether, we can safely say that in our analysis of the Roper excitation 
   the uncertainties coming from the theoretical model and approximations 
   done are
   at the level of 10-15\%.      
  
\section{Conclusion}

With the help of a theoretical model previously used to analyze the 
$(\alpha,\alpha')$ reaction on the proton, exciting the $\Delta$ in the 
projectile plus the Roper, we have analyzed data on the $(d,d')$ reaction 
on proton targets at a deuteron energy 2.3 GeV.  The use of the model 
becomes necessary because there is an important interference between the 
mechanism of delta excitation in the projectile and Roper excitation in 
the target (followed by $\pi N$ decay).   We observed that the model gave 
a good reproduction of the shape of the $\Delta$ excitation in the target, 
but the normalization exceeded the data by about $20 \%$.  In view of that 
in order to subtract this contribution and obtain the strength for Roper 
excitation, we found it justified to normalize the theoretical results by 
a factor 0.85 which leads to good agreement with the data in the $\Delta$ 
excitation region.  Similarly we multiplied by the same factor the 
interference term, calculated theoretically, and these two pieces of 
"background" were subtracted from the measured data in order to obtain the 
Roper excitation strength.  We found a qualitative agreement with the 
theoretical predictions for Roper excitation of the model, also normalized 
by the same factor.

In order to estimate uncertainties due to the use of a theoretical model 
in the analysis we compared our results to those obtained in 
Ref~\cite{rapintern}, where an empirical approach was used to subtract the 
DEP contribution using the shape of the $\Delta$ resonance excitation from 
a previous measurement at the lower energies, where there should be a 
small contribution of Roper excitation.  The results obtained with both 
methods agree qualitatively, the integrated strengths obtained are 
similar, only the peaks of the Roper strength appear a bit shifted with 
respect to each other in the two analyses.  We estimate that considering 
statistical and systematic errors, the latter ones from the model 
dependence of the subtractions, the strength of the Roper determined here 
is accurate within $25 \%$, and within these errors the agreement with 
theory can be claimed acceptable. The shape and the width of the Roper 
strength are compatible with the empirical information about the resonance.

The present analysis confirms the substantial strength for $NN \rightarrow 
NN^*$ transition in the scalar channel which has been shown already to 
have important repercussions in different physical phenomena at 
intermediate energies.

These results, and the importance that the isoscalar excitation of the 
Roper is bound to have in other process, should stimulate further 
experiments at higher energies.

\section{Acknowledgments}
We are grateful to the COE Professorship of Monbusho which enabled one of 
us, E. O., to stay at RCNP where part of this work has been done.  One of 
us, S. H., acknowledges the hospitality of SATURNE, Saclay during his 
stay and the hospitality of the University of Valencia where part of this 
work was done. This work is partly supported by DGICYT contract no. PB 96-07053 and 
an NSF grant. We are also grateful to Dr E. Tomasi-Gustafsson for very 
helpful discussions.

\pagebreak

\pagebreak

\noindent
{\bf Figure Caption}
\vskip 0.5 true cm
\noindent
Fig. 1 Diagrams for the p(d,d')X reactions considered in this paper.
They are (a) the $\Delta$ excitation in the deuteron and (b) the Roper
excitation in the proton.  The $\sigma$ exchange must be interpreted as
an effective interaction in the isoscalar exchange channel \cite{12}.

\vskip 0.5 true cm

\noindent
Fig. 2 The double differential cross section $d^2 \sigma /dM_I d\Omega$
is shown as a function of excitation energy of the proton.
The $M_I$ is the invariant mass of the target system.
Solid
circles indicate the experimental data obtained in ref. \cite{rapintern}.
The theoretical calculations are also shown in the figure, which are
total spectrum (solid line), contribution from $\Delta$ excitation
(dashed line), and contribution from Roper excitation (thick dotted line).
The Roper contributions decaying into $\pi N$ and $\pi \pi N$ are
separately shown as thin dotted lines.

\vskip 0.5 true cm

\noindent
Fig. 3 The double differential cross section $d^2 \sigma /dM_I d\Omega$
is shown as a function of excitation energy of the proton.
The $M_I$ is the invariant mass of the target system.
Solid
circles indicate the experimental data obtained in ref. \cite{rapintern}.
The theoretical calculations normalized by factor 0.85
are also shown in the figure, which are
total spectrum (solid line), contribution from $\Delta$ excitation
(dashed line), and contribution from Roper excitation (dotted line).
Solid triangles are Roper contribution extracted from the data by
subtracting the calculated $\Delta$ and interference contributions with
the normalization factor 0.85.

\vskip 0.5 true cm

\noindent
Fig. 4 The double differential cross section $d^2 \sigma /dM_I d\Omega$
is shown as a function of excitation energy of the proton.
The $M_I$ is the invariant mass of the target system.
Solid
circles indicate the experimental data obtained in ref. \cite{rapintern}.
The solid line indicates the $\Delta$ excitation contribution extracted 
from
data obtained in Ref. \cite{17} and normalized to present data at the
peak.
The Roper contribution calculated by our model and normalized by the factor
0.85 is shown by dotted line.
Solid triangles are Roper contribution extracted from the data by
subtracting the $\Delta$ contribution shown by the solid line and
the calculated interference contribution with
the normalization factor 0.85.

\vskip 0.5 true cm

\noindent
Fig. 5 The double differential cross section $d^2 \sigma /dp d\Omega$
is shown as a function of the emitted deuteron momentum at $T_d$=1.6
GeV.
Solid circles indicate the experimental data obtained in Ref. \cite{17}.
The theoretical calculations are also shown in the figure, which are
total spectrum (solid line), contribution from $\Delta$ excitation
(dashed line), and contribution from Roper excitation (dotted line).

\vskip 0.5 true cm

\noindent
Fig. 6 The double differential cross section $d^2 \sigma /dp d\Omega$
is shown as a function of the emitted deuteron momentum at
$T_d$=1.6~GeV.
Solid circles indicate the experimental data obtained in Ref. \cite{17}.
The theoretical calculations are also shown in the figure, which are
total spectrum (solid line), contribution from $\Delta$ excitation
(dashed line), and contribution from Roper excitation (dotted line).

\vskip 0.5 true cm

\noindent
Fig. 7 The double differential cross section $d^2 \sigma /dM_I d\Omega$
is shown as a function of excitation energy of the proton.
The $M_I$ is the invariant mass of the target system.
Calculated 
total spectrum (solid line) and $\Delta$ excitation contribution 
(dashed line) are shown. Thin lines are the same results as shown in Fig. 
2.  Thick lines indicate the results obtained neglecting the Fermi 
motion of nucleon in the projectile, see text.

\vskip 0.5 true cm

\noindent
Fig. 8 Same as Fig. 7 except for the normalization to the experimental 
strength.  Thin lines correspond to the calculated results in Fig. 3.


\begin{thebibliography}{99}
\bibitem{1} H.~P.~Morsch et al., Phys. Rev. Lett 69 (1992) 1336
\bibitem{2} E.~Oset, E.~Shiino and H.~Toki, Phys. Lett. B224 (1989) 249
\bibitem{3} C.~Gaarde, Nucl. Phys. A478 (1988) 475c
\bibitem{4} P.~Fern\'andez~de~C\'ordoba and E. Oset, Nucl. Phys. A544 
(1992) 793
\bibitem{5} F.~Osterfeld, Rev. Mod. Phys. 64 (1992) 491
\bibitem{6} E.~A.~Strokovsky and F.~A.~Gareev, Phys. At. Nucl. 58 (1995) 1
\bibitem{7} T.~Udagawa, S.~H.~Hong and F.~Osterfeld, Phys. Lett. B245 
(1990) 1
\bibitem{8} J.~Delorme and P.~A.~M.~Guichon, Phys. Lett. B263 (1995) 157
\bibitem{9} P.~Fern\'andez~de~C\'ordoba et al., Nucl. Phys. A586 (1995) 586
\bibitem{10} S.~A.~Coon, M.~T.~Pe\~{n}a and D.~O.~Riska, Phys. Rev. C52 
(1995) 2925
\bibitem{11} L.~Alv\'arez, E.~Oset and E.~Hern\'andez, Nucl. Phys. A633 
(1998) 519
\bibitem{12} S.~Hirenzaki, P.~Fern\'andez~de~C\'ordoba and E. Oset, Phys. 
Rev. C53 (1996) 277
\bibitem{fewbody} M.~Boivin et al, Few-Body Systems Suppl. 0 (1998) 1 and 
references therein.
\bibitem{PLB361} L.~S.~Azhgirey, E.~V.~Chernykh, A.~P.~Kobushkin, 
P.~P.~Korovin, B.~Kuehn, V.~P.~Ladygin, S.~Nedev, C.~F.~Perdrisat, 
N.~M.~Piskunov, V.~Punjabi, I.~M.~Sitnik, G.~D.~Stoletov, 
E.~A.~Strokovsky, A.~I.~Syamtomov and S.~A.~Zaporozhets, Phys. Lett. B361 
(1995) 21.
\bibitem{JINR} L.~S.~Azhgirey et al, JINR Rapid Comm 2[88] (1998) 17.
\bibitem{13} M.~P.~Rekalo and E.~Tomasi-Gustafsson, Phys. Rev. C54 (1996) 
3125
\bibitem{14} S.~Hirenzaki, E.~Oset and P.~Fern\'andez~de~C\'ordoba, Phys. 
Lett. B378 (1996) 29
\bibitem{15} S.~Hirenzaki, A.~Bacher and S.~E.~Vigdor, Phys. Rev. C59 
(1999) 1735.
\bibitem{banaigs73} J. Banaigs et al., Phys. Lett. 45B (1973) 535.
\bibitem{rapintern} C.~Djalali et al., Internal Report IPN-DRE-00/10 
(1999), Institut de Physique Nucleaire d'Orsay, France.
\bibitem{17} R.~Baldini et al., Nucl. Phys. A379 (1982) 477
\bibitem{16} R.~Machleidt, K.~Holinde and Ch.~Elster, Phys. Rep. 149 
(1987) 1
\bibitem{spes4} M.~Bedjidian et al, Nucl.~Instrum.~Method A257(1987)132.
\bibitem{pomme} B.~Bonin et al, Nucl.~Instrum.~Method A288(1990) 379.
\bibitem{norm} H.~Quechon, Ph.~D.~ thesis, University of Paris XI, Orsay 
(1980).
 \bibitem{boffi} S. Boffi, L. Bracci and P. Christillin, Nouvo Cim. A104 (1991)
 843. 
 \bibitem{carrasco} R.C. Carrasco, J. Nieves and E. Oset, Nucl. Phys. A565
 (1993) 797. 
 \bibitem{Glendenning} N. K. Glendenning and G. Kramer, Phys. Rev. 126 
 (1962) 2159.  
\end{thebibliography}
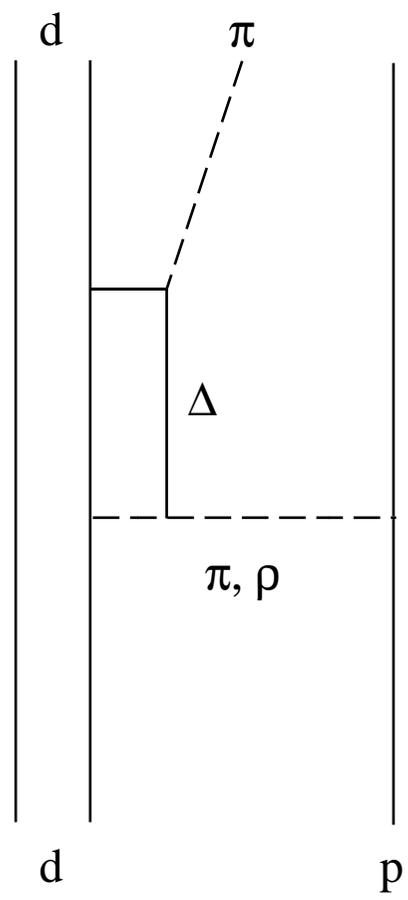
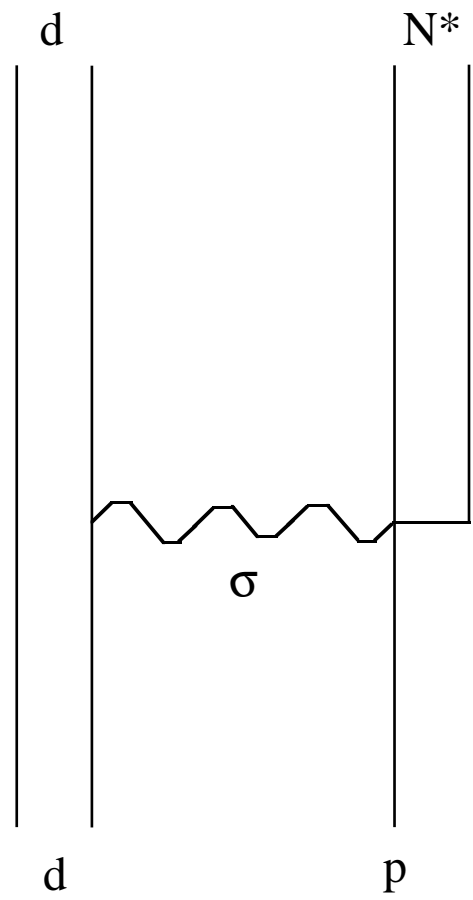
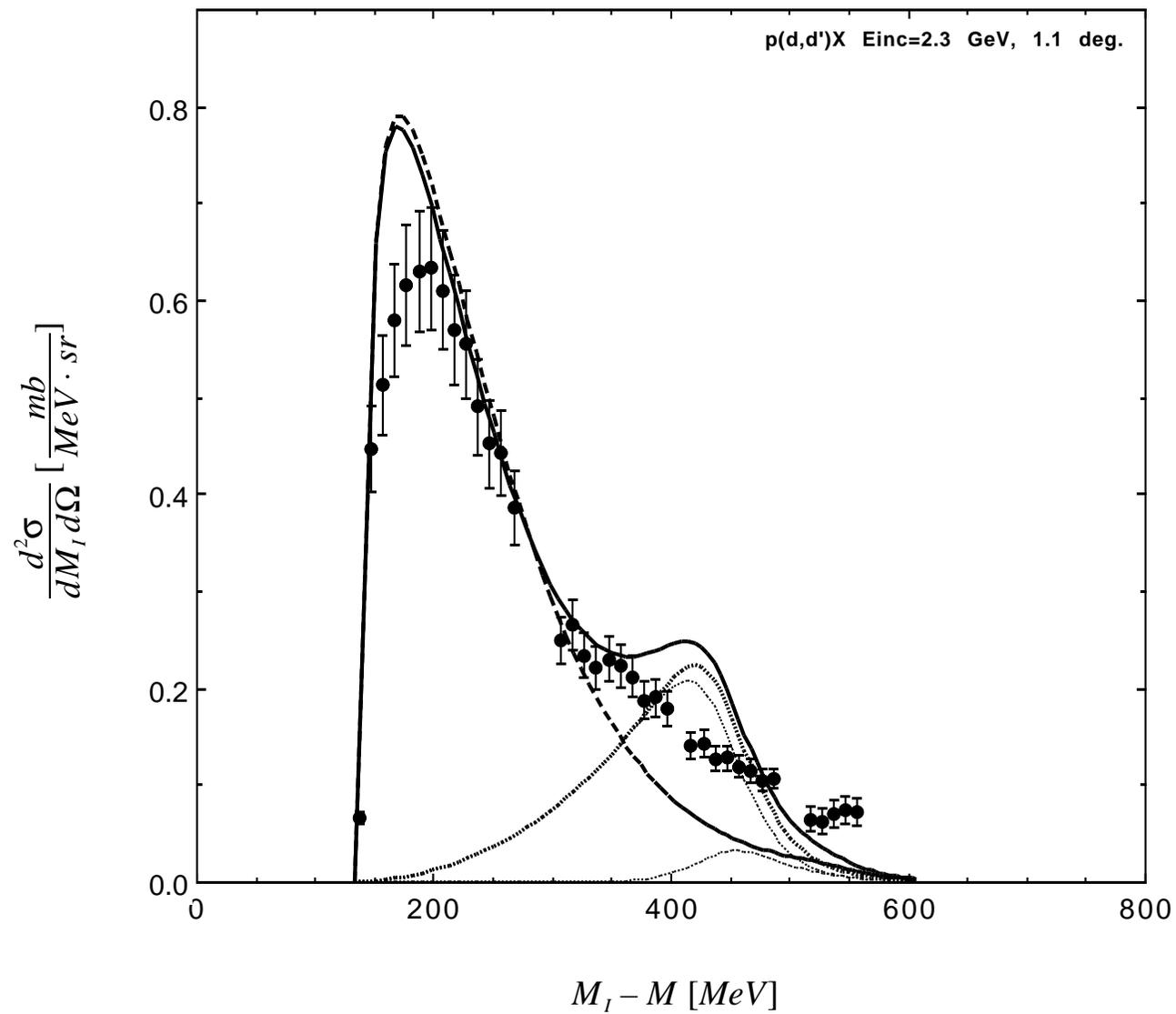
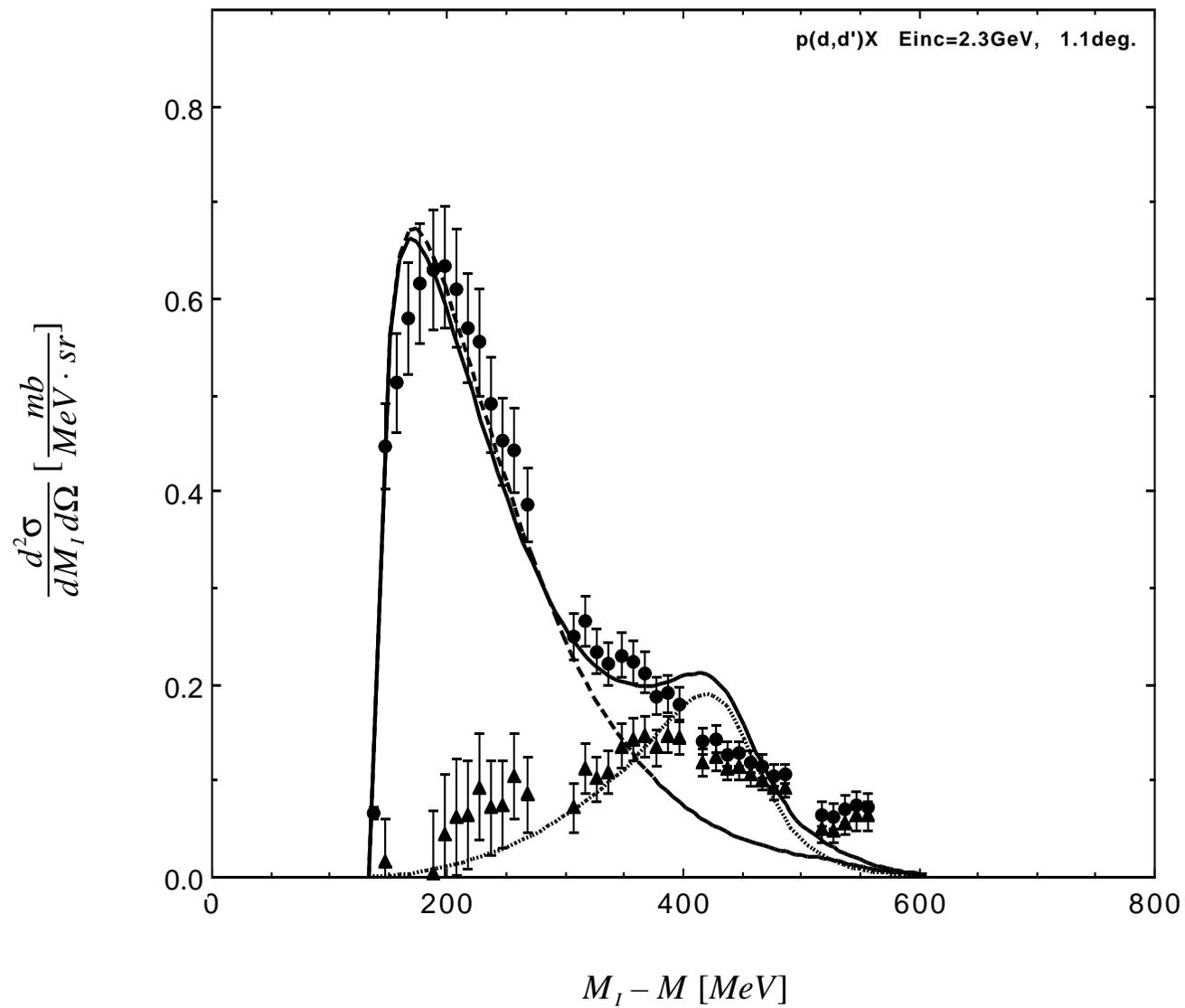
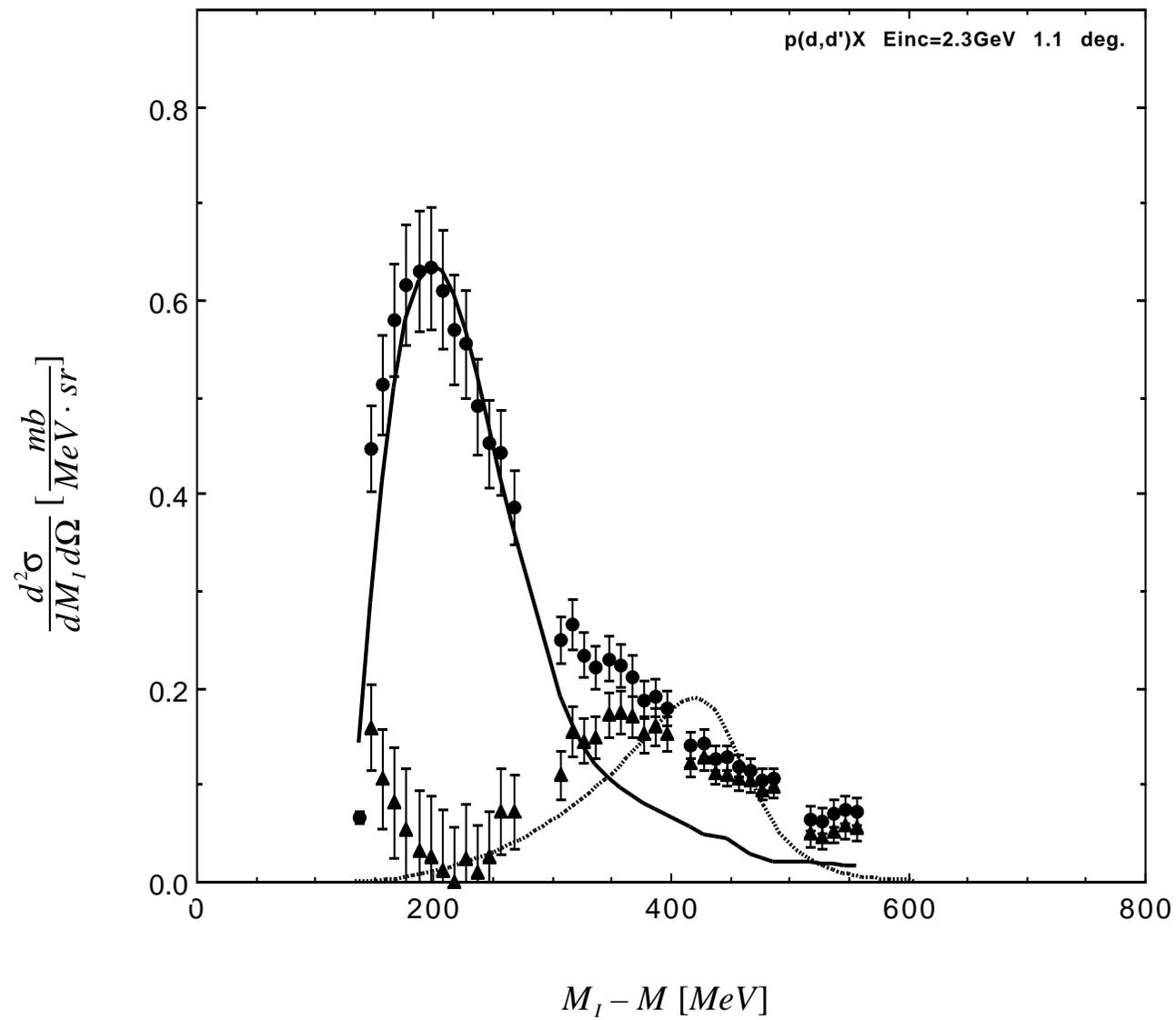
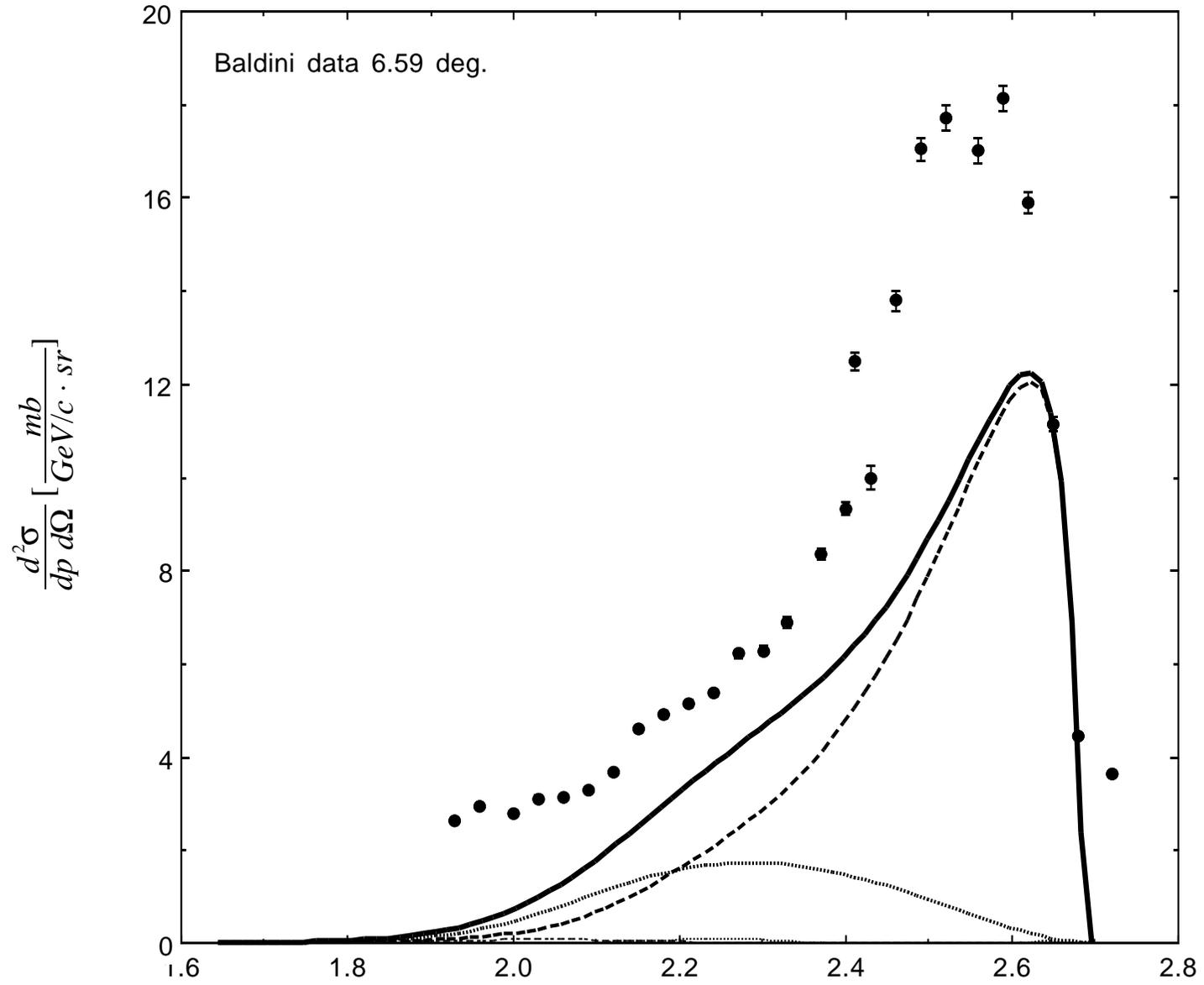
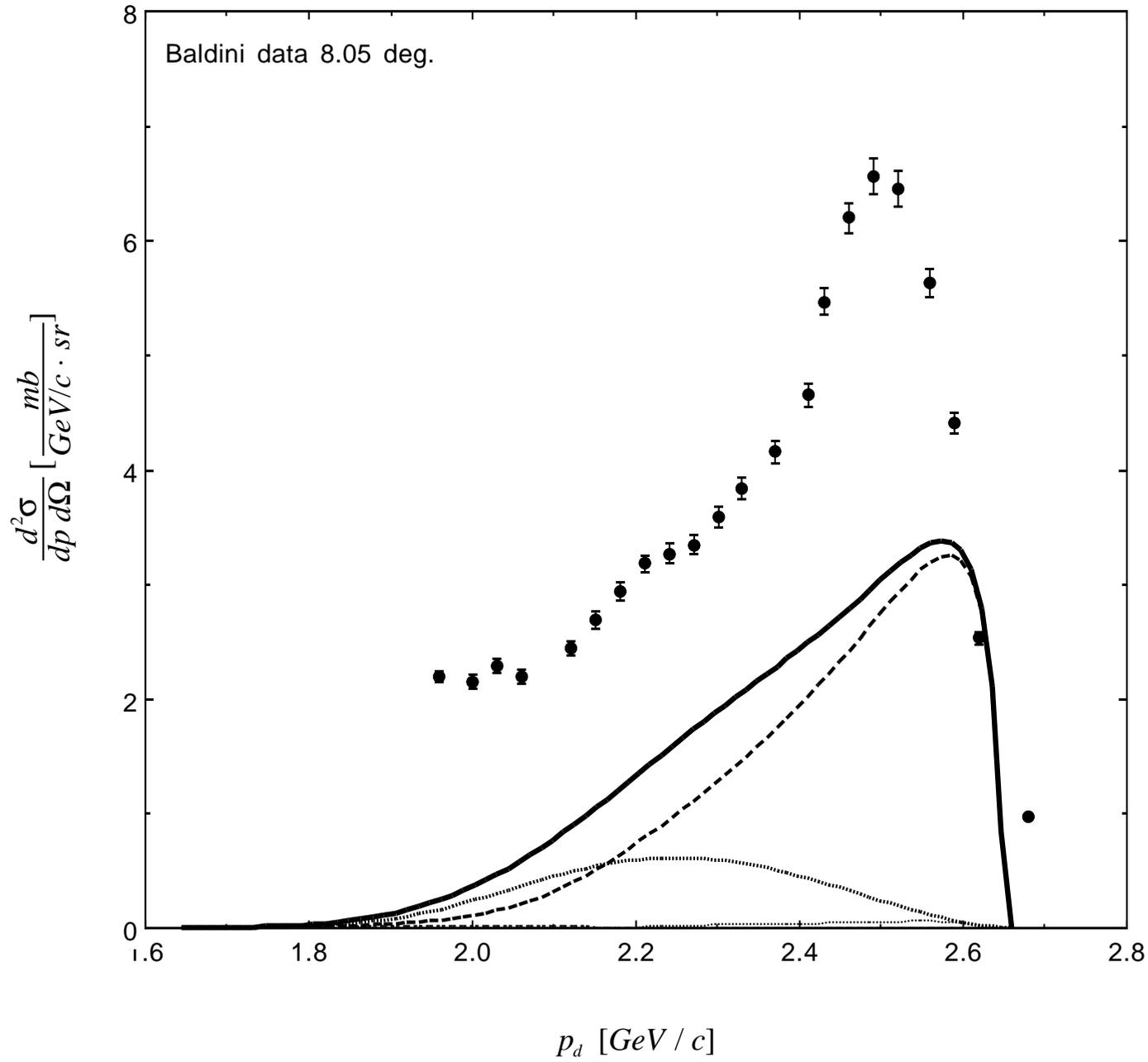
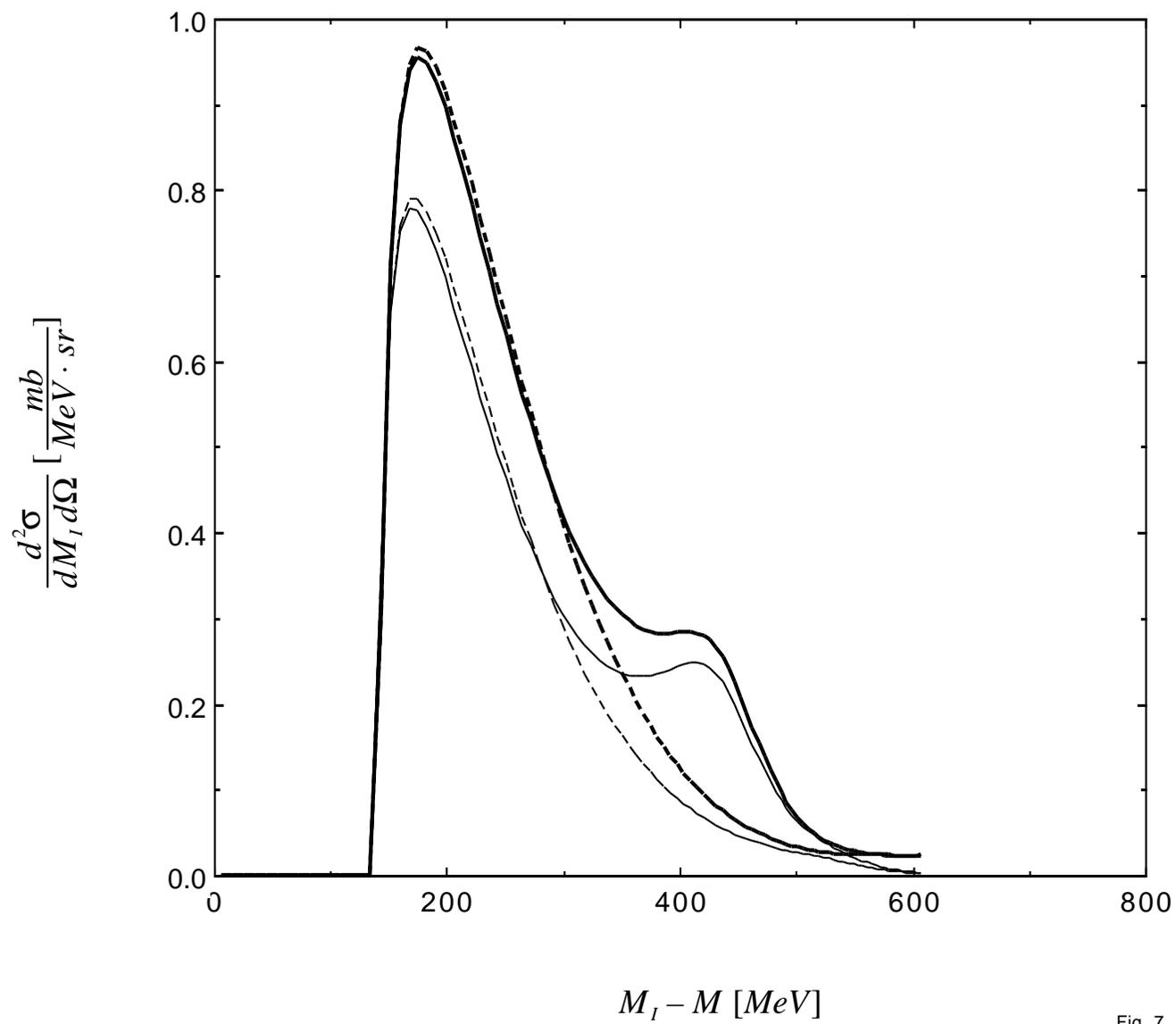
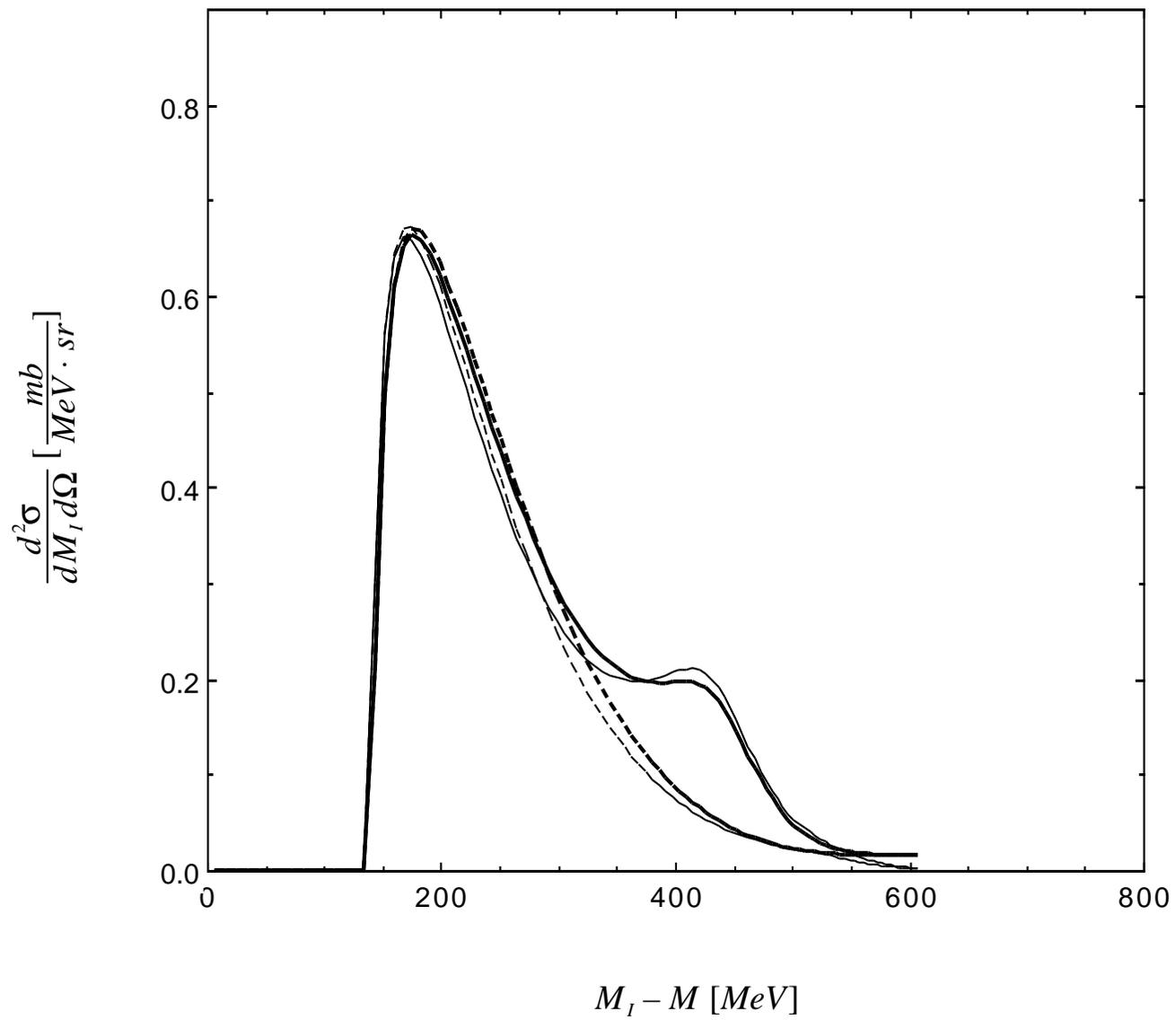
\end{document}